\documentclass[final,5p,times,twocolumn]{elsarticle}
\usepackage{amsmath}
\usepackage{amssymb}
\usepackage{bm}
\usepackage{graphicx}

\begin{document}
\begin{frontmatter}
\title{Do nonlinear waves in random media follow nonlinear diffusion equations?}

\author[mpi]{T.V.~Laptyeva\corref{cor1}}
\ead{lapteva@pks.mpg.de}
\author[mpi]{J.D.~Bodyfelt}
\author[mpi,massey]{S.~Flach}
\cortext[cor1]{Corresponding author}

\address[mpi]{Max-Planck-Institut f\"{u}r Physik Komplexer Systeme, Dresden, Germany}
\address[massey]{New Zealand Institute for Advanced Study, Massey University, Auckland, New Zealand}
             
\begin{abstract}
Probably yes, since we find a striking similarity in the spatio-temporal evolution of nonlinear diffusion equations and wave packet spreading in generic nonlinear disordered lattices, including 
self-similarity and scaling.
\end{abstract}

\end{frontmatter}

\section{Introduction}
\label{sec:intro}
The combined impact of disorder and nonlinearity strongly affects the transport properties of many physical systems leading to complex behavior contrary to their separate linear counterparts. Application has great range; particularly relevant are nonlinear effects in cold atoms \cite{Luc11, Pol10}, superconductors \cite{Dub07}, and optical lattices \cite{Per04, Sch07, Lah08}. 
Still experimental probing both disordered and nonlinear media remains limited due to reachable time or size scales. 

Significant achievements towards understanding the interplay of disorder and nonlinearity have been made in recent theoretical and numerical studies. A highly challenging problem was the dynamics of compact wave packets expanding in a disordered potential, in the
presence of nonlinearity. The majority of studies focused on two paradigmatic models - the discrete nonlinear Schr\"{o}dinger (DNLS) and the Klein-Gordon (KG) equations - revealing both destruction of an
initial packet localization and its resulting subdiffusive spreading, however with debate regarding the asymptotic spreading behaviors \cite{She93, Pik08, Fla09, Sko09, Fla10, Lap10,Bod11}. Hypotheses
of an ultimate slow-down \cite{Mul11} or eventual blockage of spreading \cite{Joh10} have been recently challenged with evidence in \cite{Iva11}, which reported a finite probability of unlimited packet expansion,
even for small nonlinearities. A qualitative theory of the nonlinear wave evolution in disordered media is based on the random phase ansatz \cite{Fla10}, derives power-law dependencies of the
diffusion coefficient on the local densities, and predicts several distinct regimes of
subdiffusion that match numerics \cite{Sko09, Fla10, Lap10, Bod11} convincingly. Closely tied to these phenomena is thermal conductivity in a disordered quartic KG chain, analyzed in \cite{Fla11}.

Similar power-law dependencies of the diffusion coefficient on the local density have been extensively studied in the context of the Nonlinear Diffusion Equation (NDE). The NDE universally describes a
diverse range of different phenomena, such as heat transfer, fluid flow or diffusion. It applies to gas flow through porous media \cite{Lei30, Mus37}, groundwater infiltration \cite{Bou03, Ame65},
or heat transfer in plasma \cite{Zel66}. As a key trait, the NDE admits self-similar solutions (also known as the source-type solution, ZKB solution or Barenblatt-Pattle solution). It describes the
diffusion from a compact initial spot and was first studied by Zel'dovich, Kompaneets, and Barenblatt \cite{Zel50, Bar52}.

The connection between nonlinear disordered spatially wave equations and NDE was conjectured recently and remains an open terrain \cite{Mul11,Ken86,Che11}. A particularly challenging question is whether the NDE self-similar
solution is an asymptotic time limit for the wave packet spreading in nonlinear disordered arrays. 
If yes, this will support the expectations that compact wave packets spread indefinitely,
without re-entering Anderson localization. In this paper, we demonstrate that the NDE captures essential features of energy/norm diffusion in
nonlinear disordered lattices. At present, we still lack a rigorous derivation of the NDE from the Hamiltonian equations for nonlinear disordered chains.
Here we show that at
sufficiently large time the properties of the NDE self-similar solution reasonably approximate those
of the energy/norm density distribution of nonlinear waves; manifesting in similar asymptotical behaviors of statistical measures (such as distribution 
moments and kurtosis), and in the overall scaling of the density profiles. To substantiate our conclusions, we perform simulations of a modified NDE and compare the results against the spatio-temporal evolution of nonlinear disordered
media \cite{Lap10, Bod11}. 
\section{Theoretical predictions}
\subsection{Basic nonlinear disordered models}
\label{sec:models}
The spreading of wave packets has been extensively studied within the framework of KG/DNLS arrays. Particularly, the DNLS describes the wave dynamics in various experimental
contexts, from optical wave-guides \cite{Sch07, Lah08} to Bose-Einstein condensates \cite{Sme03}. The DNLS equation approximates well the KG one under appropriate conditions of small norm densities
\cite{Kiv92, Joh06}. The KG has the advantage of faster integration at the same level of accuracy.

The DNLS chain is described by the equations of motion
\begin{equation}
i \dot{\psi_l} = \epsilon_l \psi_l + \beta \left| \psi_l \right|^2 \psi_l - \psi_{l+1} - \psi_{l-1}, \label{eq:DNLS}
\end{equation}
where $\epsilon_l$ is the potential strength on the site $l$, chosen uniformly from an uncorrelated random distribution $\left[ -W/2,W/2 \right]$ parameterized by the disorder strength $W$.

The KG lattice is determined by
\begin{equation}
\ddot{u}_l  = -{\tilde \epsilon}_l u_l - u_l^3 + \frac{1}{W}(u_{l+1} + u_{l-1} - 2u_l), \label{eq:KG}
\end{equation}
where $u_l$ and $p_l$ are, respectively, the generalized coordinate/momentum on the site $l$ with an energy density $E_l$. The disordered potential strengths $\tilde{\epsilon_l}$ are chosen uniformly
from the random distribution $[1/2,3/2]$. The total energy $\bar{E} = \sum_l E_l$ acts as the nonlinear control parameter, analogous to $\beta$ in DNLS (see e.g. \cite{Sko09}). 

Both models conserve the total energy, the DNLS additionally conserves the total norm $\mathcal{S} = \sum_{l} |\psi_l|^2$. The approximate mapping from the KG to the DNLS is $\beta \mathcal{S} \approx
3W\bar{E}$. We restrict analytics to the DNLS model, since the results are transferable to the KG one.
\subsubsection{Spreading predictions}
\label{sec:an_pred}
In order to quantitatively characterize the wave-packet spreading in Eqs.(\ref{eq:DNLS},\ref{eq:KG}) and compare the outcome to the NDE model, we track the probability at the $l$-th site,
$\mathcal{P}_l \equiv n_l = |\psi_l|^2$, where $n_l$ is the norm density distribution. Note that the analog of $n_l$ in the KG is the normalized energy density distribution $E_l$. We then track a
normalized probability density distribution, $z_l \equiv n_l / \sum_k n_k$.  In order to probe the spreading, we compute the time-dependent moments $m_{\eta} = \sum_l z_l (l - \bar{z})^\eta$,
where $\bar{z} = \sum_l l \: z_l$ gives the distribution center. 

We further use as an additional dynamical measure the kurtosis \cite{Dod03}, defined as
$\gamma(t)=m_4(t)/m^2_2(t)-3$. Kurtosis is an indicator of the overall shape of the probability distribution profile - in particular, as a deviation measure from the normal profile. Large
values correspond to profiles with sharp peaks and long extending tails. Low values are obtained for profiles with rounded/flattened peaks and steeper tails. As an example, the Laplace distribution
has $\gamma = 3$, while a compact uniform distribution has $\gamma = -1.2$.

The time dependence of the second moment $m_2$ of the above distributions $z_l$ was previously derived and studied in \cite{Fla09, Sko09, Fla10, Lap10, Bod11}. Different regimes of energy/norm
subdiffusion were observed and explained. Generally, $m_2$ follows a power-law $t^{\alpha}$ with $\alpha<1$. Here we briefly recall the key arguments.

In the linear limit Eqs.(\ref{eq:DNLS},\ref{eq:KG}) reduce to the same eigenvalue problem \cite{Fla09, Sko09}. We can thus determine the normalized eigenvectors $A_{\nu,l}$ and the eigenvalues
$\lambda_{\nu}$. With $\psi_l = \sum_{\nu} A_{\nu,l}\phi_{\nu}$, Eq.(\ref{eq:DNLS}) reads in an eigenstate basis as 
\begin{equation}
i\dot{\phi}_{\nu}=\lambda_{\nu}\phi_{\nu} + \beta \sum_{\nu_1,\nu_2,\nu_3} I_{\nu,\nu_1,\nu_2,\nu_3} \phi^{\ast}_{\nu_1} \phi_{\nu_2}\phi_{\nu_3}, \label{eq:NME}
\end{equation}
where $I_{\nu,\nu_1,\nu_2,\nu_3}=\sum_l A_{\nu,l}A_{\nu_1,l}A_{\nu_2,l}A_{\nu_3,l}$ are  overlap integrals and $\phi_{\nu}$ determine the complex time-dependent amplitudes of the
eigenstates.

In \cite{Fla10} the incoherent ``heating'' of cold exterior by the packet has been established as the most probable mechanism of spreading. Following this analysis, the packet modes $\phi_{\nu}(t)$ should
evolve chaotically  with a continuous frequency spectrum. In particular chaotic dynamics of phases
is expected to destroy localization. The degree of chaos is linked 
to the number of resonances, whose probability
becomes an essential measure for the spreading. Previous studies \cite{Kri10} indicate that the probability of a packet eigenstate to be resonant is $\mathcal{R}(\beta n) = 1-e^{-C
\beta n}$, with $C$ being a constant dependent on the strength of disorder. 
The heating of an exterior mode close to the edge of the wave packet with norm density $n$
would then follow
\begin{equation}
i\dot{\phi}_{\mu} = \lambda_{\mu}\phi_{\mu} + \beta n^{3/2} \mathcal{R}(\beta n) f(t)
\label{eq:Langevin}
\end{equation}
with delta-correlated (or reasonably short-time correlated) noise $f(t)$, and lead to
$\left|\phi_{\mu}\right|^2 \sim \beta^2 n^3 (\mathcal{R}(\beta n))^2 t$. The momentary diffusion rate 
follows as $D \sim \beta^2 n^2 (\mathcal{R}(\beta n))^2$.

With $m_2 \sim n^{-2} = Dt$ one arrives at $1/n^2 \sim \beta (1-e^{-C \beta n}) t^{1/2}$. Depending on the product $C \beta n$ being larger or smaller than one, the packet has two regimes
of subdiffusion (and a dynamical crossover between them): $m_2 \sim \beta t^{1/2}$
(strong chaos) and $m_2 \sim \beta^{4/3} t^{1/3}$ (asymptotic weak chaos) 
\cite{Fla09, Sko09, Fla10, Lap10, Bod11}.

The validity of the assumption of incoherent phases and of Eq.(\ref{eq:Langevin}) was established through
numerical studies for the first time by Michaely and Fishman \cite{Mich12}, moving the above
conjecture based theories onto solid grounds.

\subsection{Nonlinear diffusion equation}
\label{sec:nde}
The assumption of chaos and random phases, Eq.(\ref{eq:Langevin}), the density dependent diffusion
coefficient and the resulting subdiffusion
strongly suggest an analogy to nonlinear diffusion equations (see e.g. Ref.\cite{Kol10}).
We first consider here the much studied NDE with a power-law dependence of the diffusion coefficient on the density. The NDE in the one-dimensional case reads
\begin{equation}
\partial_t \mathcal{P} = \partial_x \left(\kappa \mathcal{P}^a \partial_x \mathcal{P} \right) \label{eq:NDE}
\end{equation}
Here $\mathcal{P} \equiv \mathcal{P}(x,t)$ is the concentration of the diffusing species 
(which may be related to the energy/norm density), $a > 0$ and $\kappa$ are some constants. Hereafter, we set $\kappa=1$ without loss of generality.

Let us discuss the scaling properties of Eq.(\ref{eq:NDE}). Given a solution $\mathcal{P}(x,t)$ it
follows that
\begin{equation}
\mathcal{P}(x,t) = s_p \mathcal{P}(s_x x,s_t t) \label{ndescaling}
\end{equation}
where $s_p,s_x,s_t$ are some scaling factors. From normalization it follows that
$s_p=s_x$. Inserting (\ref{ndescaling}) into (\ref{eq:NDE}) we find
\begin{equation}
s_t=s_x^{a+2}\;. 
\label{ndescales}
\end{equation}
After some algebra for the moment $m_{\eta}(t)$ we conclude
\begin{equation}
m_{\eta}=\left( \frac{t}{t_0}\right) ^{\eta/(a+2)} m_{\eta}(t_0) 
\label{ndemomentscaling}
\end{equation}
which corresponds to a subdiffusive process.

\begin{figure*}
\begin{center}
\includegraphics[width=1.8\columnwidth,keepaspectratio,clip]{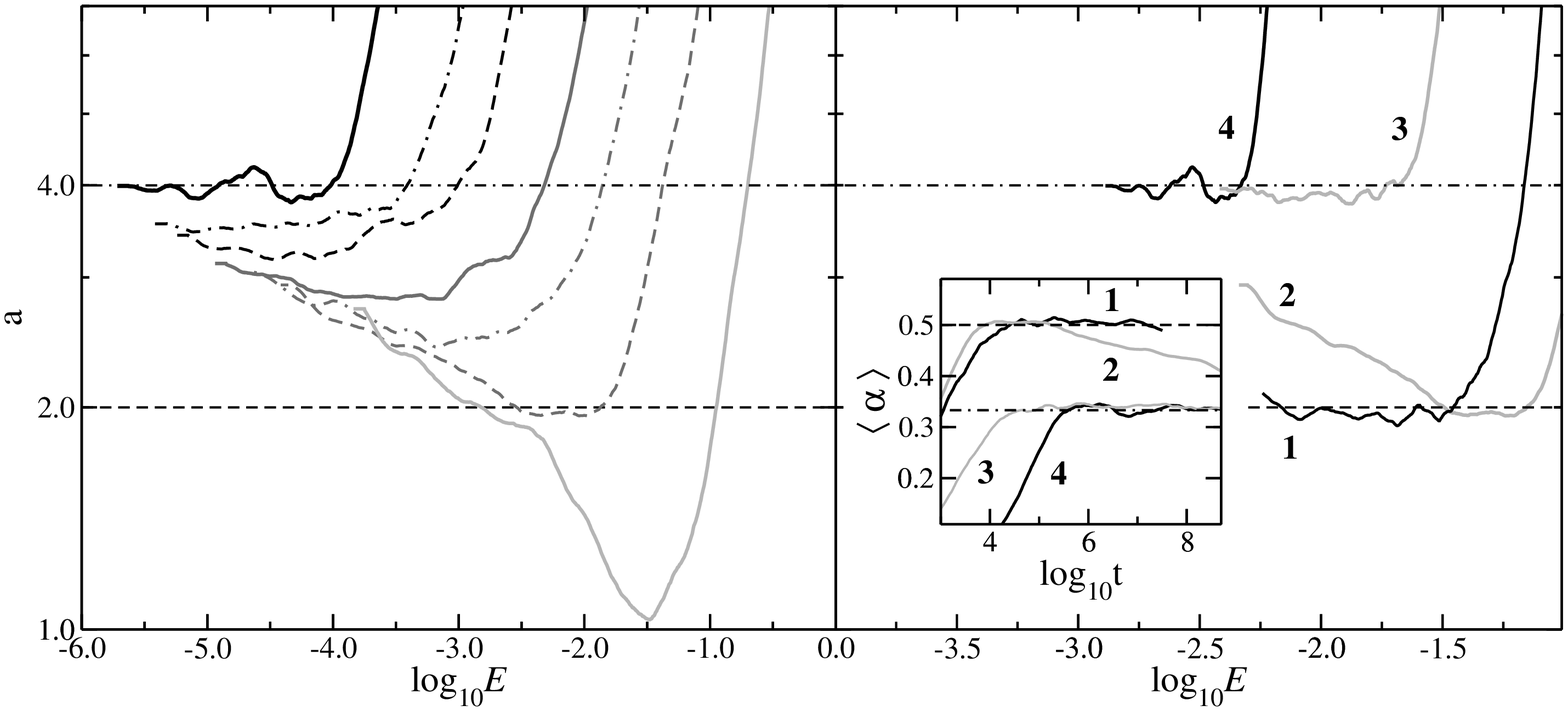}
\caption{The NDE parameter $a$ versus the energy density $E(t) \sim E(0)\left\langle m_2(t) \right\rangle^{-1/2}$ as obtained from numerical simulations of the KG chain. 
Horizontal lines guide the eye at values $a=2$ and $a=4$.
\textit{Left panel}: $W=4$, $L=21$, and $E \in \left\{0.01, 0.015, 0.02, 0.04, 0.06, 0.1, 0.75\right\}$ (varying from the top to bottom and from weak chaos to strong chaos and self trapping).
\textit{Right panel}: $W=2$ and $E=0.1$ (curve 1, strong chaos), $W=4$ and $E=0.2$ (curve 2, crossover from the strong to weak chaos), $W=4$ and $E=0.01$ (curve 3, weak chaos), $W=6$ and $E=0.05$
(curve 4, weak chaos). 
\textit{Inset}: Dependence of $\alpha$ on time for the same data as in the right panel.
The dot-dashed and dashed lines correspond to the values $\alpha =1/3$
and $\alpha=1/2$ respectively.}
\label{fig2}
\end{center}
\end{figure*}

Eq.(\ref{eq:NDE}) has various self-similar solutions \cite{Vaz07, Pol04, Ame65} 
which depend on the characteristics of the evolving state. For compact wave packets
the self-similar state realized in the asymptotic limit of large time is of the form
\cite{Zel50, Bar52, Tuc76}
\begin{equation}
\mathcal{P}(x,t) = \left[ A - \dfrac{a x^2}{2(2+a)t^{2/(2+a)}} \right]^{1/a} t^{-1/(2+a)}, \label{eq:SSS}
\end{equation}
where $A$ is a normalization constant and $|x| < x_0$. For $|x| > x_0$ the density $\mathcal{P}$ strictly
vanishes. The edge position $x_0$ depends on time as
\begin{equation}
x_0 = \left[ 2 A (2/a+1) t^{2/(2+a)} \right]^{1/2}. \label{eq:x0}
\end{equation}
The linear stability of Eq.(\ref{eq:SSS}) was demonstrated in
\cite{Wit98, Mat06}.

Using a change of variables $y = x/x_0$ we obtain for the density $\bar{\mathcal{P}}(y,t)$
with $\bar{\mathcal{P}} {\rm d}y = \mathcal{P} {\rm d}x$:
\begin{equation}
\bar{\mathcal{P}}(y,t) = A^{1/a+1/2} \sqrt{4/a+2} (1-y^2)^{1/a}\;. \label{ndesolution}
\end{equation}
Since $\bar{\mathcal{P}}$ does not depend on time, it follows
\begin{equation}
m_{\eta}(t)=x_0^{\eta} \bar{m}_{\eta}\;,\;\bar{m}_{\eta} = \int_{-1}^{1}y^{\eta} \bar{\mathcal{P}} dy
\;. \label{ndem2}
\end{equation}
In agreement with (\ref{ndemomentscaling})
this yields e.g. $m_2 \sim t^{2/(2+a)}$.

The moments of solution (\ref{eq:SSS})
\begin{equation}
m_{\eta} =
 \left[ \dfrac{2(2+a)}{a} \right]^{\frac{\eta+1}{2}}
\mathcal{B}\left[\frac{a+1}{a};\frac{\eta+1}{2}\right] A^{\frac{a(\eta+1)+2}{2a}} t^{\frac{\eta}{2+a}}
\;, \label{moments}
\end{equation}
where $\mathcal{B}(x,y)$ is the Euler Beta Function.
Using Eq.(\ref{moments}) we derive for the kurtosis
\begin{equation}
\gamma_\infty \equiv \lim_{t\rightarrow \infty}\gamma=3\:\frac{\Gamma\left[\frac{5}{2}+\frac{1}{a}\right] \Gamma\left[ \frac{5}{2}+\frac{1}{a}\right]}
{\Gamma\left[ \frac{3}{2}+\frac{1}{a}\right]\Gamma\left[\frac{7}{2}+\frac{1}{a}\right]}-3, \label{eq:kurtosis}
\end{equation}
where $\Gamma[x]$ is Legendre's Gamma function. As can be seen, the kurtosis of the self-similar solution does not depend on time. For the values $a=4$
and $a=2$ we obtain kurtosis values $\gamma=-1.091$ and $\gamma=-1.00$ respectively. Additionally, $\gamma = -1.20$ in the limit of $a \rightarrow \infty$, which corresponds to a flat uniform
distribution. 

Few remarks are in place in order to proceed. First, the spatial discretization of the NDE
(\ref{eq:NDE}) by introducing discrete Laplacians with nearest neighbor differences does not modify
the properties of the asymptotic states \cite{Kol10}. However, the overlap integrals in (\ref{eq:NME})
decay exponentially in space. Moreover, the diffusion coefficient $D$ for spreading wave packets
is in general different from a pure power of the density (see (\ref{eq:Langevin}) and below).
It takes a power function only in the asymptotic regime of weak chaos, and the potential long lasting
intermediate strong chaos regime. Therefore we will generalize and adapt the above NDE in the next subsection.
\subsection{Modified nonlinear diffusion equation}
\label{sec:recnde}
Firstly, we 
rewrite Eq.(\ref{eq:NDE}) as
\begin{equation}
\partial_t \mathcal{P} = \frac{1}{a+1} \partial^2_x \mathcal{P}^{a+1}. \label{eq:NDE_num}
\end{equation}
Since density leakage into neighboring sites is directly related to resonance probability \cite{Fla09,
Sko09, Fla10, Lap10, Bod11}, we introduce it into the RHS of Eq.(\ref{eq:NDE_num}) as
\begin{equation}
\mathcal{P}^{a+1} \rightarrow F = \mathcal{P}^{a+1} \left(1 - e^{-C \mathcal{P}}\right)^2 \nonumber
\end{equation}
Randomness in disordered systems exponentially localizes the normal modes, so mode-mode coupling in the nonlinear overlap integral has an exponential dependence in distance as well. 
We therefore introduce an exponentially decaying interaction along a discrete chain. Using a finite central difference, we arrive at the modified nonlinear diffusion equation (MNDE)
\begin{equation}
\partial_t \mathcal{P}_n = \sum_{m>n} e^{-m/\chi} \left( F_{n-m} -2 F_n + F_{n+m} \right). \label{eq:NDE_Exp}
\end{equation}
In the above, we treat $C$ as a free parameter, and equal $\chi$ in our numerics to localization lengths of the disordered lattice models. We fix $a=2$ for the MNDE, and 
expect to observe similar regimes, such that for $\mathcal{P}_l \sim 1$, we have weak chaos for $C\ll 1$ and strong chaos in the opposite limit of $C\gg 1$. The above MNDE is therefore expected to
account for the resonance probabilities between normal modes,
and the exponentially decaying interaction between them.

\subsection{Numerical simulations of MNDE}
\label{sec:recnde_sim}
We integrate Eq.(\ref{eq:NDE_Exp}) with a fourth-order
Runge-Kutta scheme \cite{Pre96}, for a number of values for the free parameter $C$. We start
with an initially compact distribution of width $L=41$ and density
$\mathcal{P}_{\left|l \right| \leq L/2}=1$ (hereafter referred to as \textit{brick} distribution) and 
$\mathcal{P}_{\left|l \right| > L/2}=0$. The integrations were carried out to
$t \approx 10^6$ using a time step of $0.4$, all the while conserving norm to
better than $10^{-12}$. 

\begin{figure}[h]
\begin{center}
\includegraphics[width=0.9\columnwidth,keepaspectratio,clip]{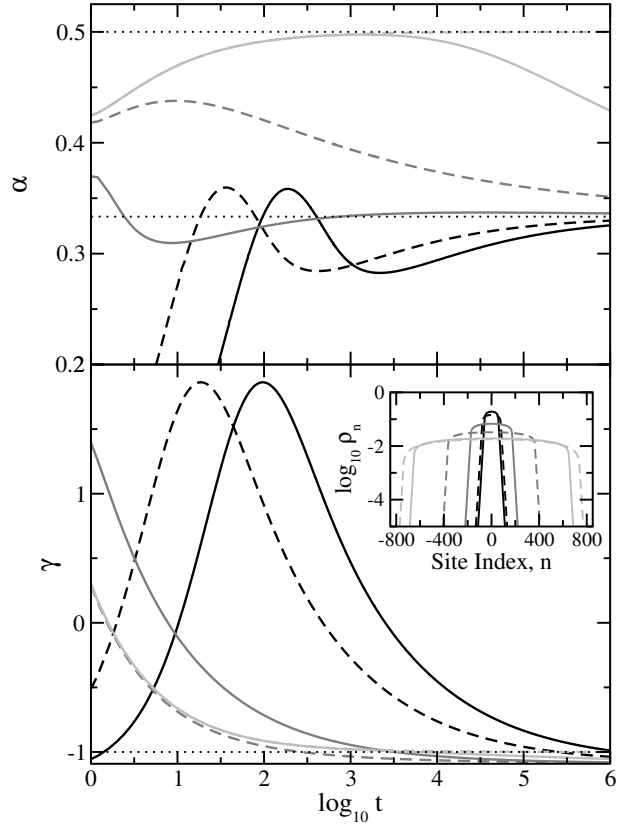}
\caption{Numerical results for the MNDE, Eq.(\ref{eq:NDE_Exp}) with $\chi=6$ and $a=2$
and initial conditions given in the text.
\textit{Upper:} The exponent $\alpha(t)$ for the MNDE with the following values of $C$:
$0.1$ (dashed black), $1.0$ (solid black), $10$ (dashed dark grey), $25$ (solid dark
grey), $50$ (dashed light grey), and $1000$ (solid light grey).
\textit{Lower:} The kurtosis $\gamma(t)$ for the same simulations.
Inset: The density profiles of the above simulations 
at $t=10^6$.
}
\label{fig1}
\end{center}
\end{figure} 

From the second moment, $m_2(t)$, we compute the derivative 
\begin{equation}
\alpha (t) = \frac{{\rm d}\log_{10} m_2 }{ {\rm d} \log_{10} t } 
\label{exponent}
\end{equation}
and plot the result in the upper panel of  Fig.\ref{fig1}.
We find that the MNDE reproduces weak
chaos ($\alpha = 1/3$ for $C \leq 1 $) and intermediate strong chaos ($\alpha =
1/2$  for $C \gg 1$). 

We further plot in Fig.\ref{fig1} the time-dependent kurtosis (lower panel) and
the density distributions at $t=10^6$ (inset) for a few representative values of $C$. 
Note that those
states that are in the weak chaos regime ($C\lesssim 1$) show a tendency
toward an asymptotic $\gamma_\infty \approx -1.091$ in agreement with the NDE, 
but not reaching it fully in our
simulations. Those states in the intermediate strong chaos regime ($C \gtrsim 100$) exhibit
long lasting saturation at $\gamma_\infty \approx -1$ in agreement with the NDE. 
We also observe a growth of the kurtosis into positive values for weak chaos,
followed by a drop that decays to $-1.091$.

\section{Wave packet spreading in nonlinear disordered lattices}
\label{sec:num_disor}
Let us discuss first the details of computations. For both models of Eqs.(\ref{eq:DNLS},\ref{eq:KG}) we consider initial brick distributions of width $L$ with nonzero internal energy
density (or norm density for the DNLS) and zero outside this interval. In contrast to the NDE
and MNDE, these lattice systems, in addition to local densities, 
are also characterized by local phases. Initially the phase at each site is set randomly.
Equations (\ref{eq:DNLS}), (\ref{eq:KG}) are evolved using SABA-class
split-step symplectic integration schemes \cite{LR01}, with time-steps of $10^{-2} - 10^{-1}$.  
Energy conservation is within a relative tolerance of less than $0.1\%$. We perform ensemble averaging over $10^3$ realizations of the onsite disorder.

With $m_2 \sim t^{2/(2+a)}$ of the NDE self-similar solution, Eq.(\ref{moments}), and $m_2 \sim t^{\alpha}$ for KG/DNLS models, the NDE parameter $a$ is related to the
exponent $\alpha$ as $a=2(1-\alpha)/\alpha$. This allows a monitoring of
$a$ as the energy density changes. We expect then $a=2$ and $a=4$ respectively for the
strong and weak chaos regimes, as well as a shift from $a=2$ to $a=4$ associated with the  
crossover between the two regimes. 
\begin{figure}[h]
\begin{center}
\includegraphics[width=0.9\columnwidth,keepaspectratio,clip]{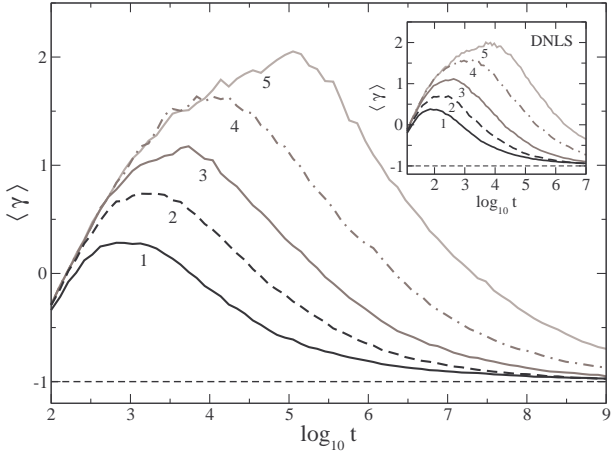}
\caption{Semi-log plot of average over $10^3$ realizations kurtosis $\left\langle\gamma\right\rangle$ versus time for KG (see main part) and DNLS (see inset) models with parameters $W=4$, $L=21$.
Numbers correspond to different energy densities (KG), or, nonlinearity strengths (DNLS): $E=0.01$ or $\beta=0.04$ (curve 1), $E=0.02$ or $\beta=0.08$ (curve 2), $E=0.04$ or $\beta=0.18$ (curve 3),
$E=0.08$ or $\beta=0.36$ (curve 4), $E=0.2$ or $\beta=0.72$ (curve 5). The dashed lines correspond to the $\left\langle \gamma \right\rangle=-1.0$.}
\label{fig3}
\end{center}
\end{figure}

We validate these predictions using our numerical data that correspond to the different regimes of spreading \cite{Lap10}. We plot the NDE parameter $a$ from the numerically obtained $\alpha$ (see
inset in Fig.\ref{fig2}), assuming that the energy density $E(t) \sim E(0)\left\langle m_2(t)\right\rangle^{-1/2}$ in Fig.\ref{fig2}. As predicted, $a$
reaches the asymptotic value $a=4$ for weak chaos. Our numerical results also show that 
once $a$ reaches its asymptotic value $a=4$, it does not increase further in time,
even for quite small energy densities. That is a clear indication for the absence of speculated
slowing down dynamics \cite{Mul11}. For strong chaos $a$ temporarily saturates around
$a=2$, keeps this value only within some interval of energy densities, and finally crosses over into the interval $2<a<4$ with a clear tendency to reach the weak chaos value $a=4$ at larger times.

The resulting kurtosis evolution, $\left\langle \gamma(t) \right\rangle$ is presented in Fig.\ref{fig3}. For the initial wave packet $\left\langle \gamma(0)
\right\rangle = -1.2$. The kurtosis first displays a transient increase to positive values. This is very similar to the MNDE results and is due to
exponential localization of the initial state in normal mode space. 
At larger times
$\left\langle \gamma(t) \right\rangle$ displays a decrease in time, approaching the self-similar behaviors in density distributions with $\left\langle \gamma \right\rangle \approx -1$ (recall that
the NDE self-similar solution gives us $\gamma = -1$ for $a=2$ and $\gamma = -1.091$ for $a=4$). 
\begin{figure}[h]
\begin{center}
\includegraphics[width=0.9\columnwidth,keepaspectratio,clip]{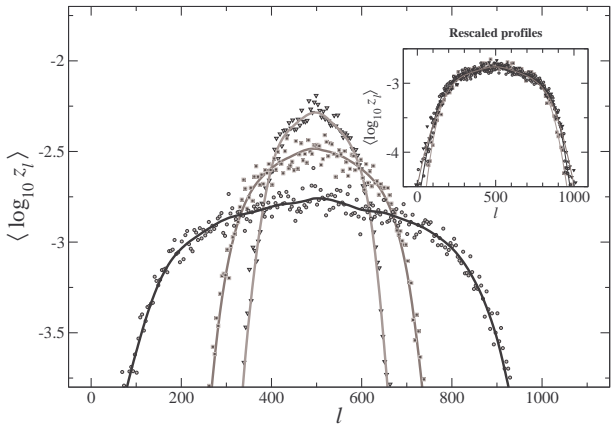}
\caption{\textit{Main:} The log of the normalized energy density distribution $\left\langle \log_{10}z_l \right\rangle$ at three different times (from the top to bottom 
$t \approx 10^4$, $t \approx 10^7$, $t \approx
10^8$). The initial parameters are $E=0.2$, $W=4$ and $L=21$. Symbols correspond to 
the average over $10^3$ disorder realizations, and solid lines correspond to an additional smoothing. 
\textit{Inset:} Rescaled distributions (see text).}
\label{fig4}
\end{center}
\end{figure}

The evolution of the averaged energy density profiles $\left\langle E \right\rangle$ in the course of spreading is illustrated in Fig.\ref{fig4}. The peaked initial
distribution profiles transform into more flat ones as time evolves. 
The most striking result is obtained by rescaling the profiles in Fig.\ref{fig4} according to
the scaling laws of the NDE. We estimate the value of $x_0$ (see (\ref{eq:x0})) as the 
position where the profiles in Fig.\ref{fig4} reach $10^{-4}$. We then plot the rescaled
densities according to (\ref{ndesolution}) in the inset of Fig.\ref{fig4}. We observe very good
scaling behavior. This is the strongest argument to support the applicability of NDE and MNDE
to the spreading of wave packets in nonlinear disordered systems. It also strongly supports
that the spreading process follows the predicted asymptotics and does not slow down or even halt.

\section{Conclusion}
We scrutinized the suggested connection between the temporal evolution of self-similar solutions of the NDE and MNDE on one side, and the asymptotic dynamics of the energy/norm distribution within nonlinear disordered media on the other, and found a remarkable correspondence. In order to describe the expansion of an initial distribution with time, we used two key quantities: (i) the second moment $m_2(t)$, which shows how the squared width of the distribution grows; (ii) the kurtosis $\gamma(t)$, which indicates how the shape of the distribution profile changes. 

As a first test we compared the exponents characterizing the subdiffusion for the second moments of the energy/norm density distributions. In KG/DNLS models, $m_2(t) \sim t^\alpha$, and for the NDE self-similar solution, $m_2(t) \sim t^{2/(2+a)}$, the exact identity giving $\alpha = 2/(2+a)$. We found that the wave packets in nonlinear disordered chains converge towards the self-similar behavior at large times. The numerical results show a good correspondence to the NDE-based analytics in a wide range of parameters.

Second, in disordered lattices the energy/norm distributions have exponentially decaying tails at variance to the steep-edged NDE self-similar solution. Such difference of energy/norm distribution profiles has no effect on $m_2(t)$  at large times. However, it leads to differences between the 
NDE and the KG/DNLS dynamics at intermediate times, e.g. seen in the temporal behavior of 
the kurtosis $\left\langle\gamma(t)\right\rangle$. 
In order to study the possible impact of various initial density profiles, we also computed the time evolution of the NDE for initial probability distributions with exponential tails. In all simulations with NDE parameter $a \leq 6$, the kurtosis asymptotically reached the expected value $\gamma_\infty$ being a function of the parameter $a$ only.

To bridge a possible  gap between the NDE and the disordered nonlinear lattices, we introduced
a modified MNDE. To account for the interaction between localized Anderson lattice modes we implemented 
exponentially decaying coupling in the MNDE, and also incorporated the resonance probability of
normal modes into a modified nonlinear diffusion coefficient. 
Then we indeed observe the dynamical behavior that reproduces the spatio-temporal evolution of nonlinear disordered chains, namely the weak and the strong chaos regimes of spreading, 
and he correct temporal evolution of the kurtosis, and the distribution profiles. 
Most importantly we observe the precise scaling behavior of the asymptotic NDE solutions in the 
case of nonlinear wave packets.

Let us summarize. There is a lack of knowledge on the statistical properties of chaotic dynamics generated by nonlinear coupling. We are still far from a rigorous derivation of the NDE when
starting with the equations of motion for nonlinear disordered chains. Nevertheless, the theory for initial energy/norm spreading in KG/DNLS chains which is based on a Langevin dynamics approximation has earlier been confirmed by exhaustive numerical studies. Of course, there is difference between KG/DNLS nonlinear disordered models and the NDE. Despite this, numerical results confirm that at sufficiently large time the NDE self-similar solution approximates remarkably well the spreading properties of energy/norm density distributions in terms of the second moment, the kurtosis, and the scaling features. Therefore, the NDE as a simple analytical tool is extremely useful for studying the initial excitation spreading in nonlinear disordered media at asymptotically large times. Additionally, the NDE analog with long-range exponentially decaying coupling shows an even deeper correspondence between generic nonlinear disordered models and the NDE and therefore might prove to be an insightful model  for the future analysis of spreading in nonlinear disordered systems.

We thank N.~Li, M.V.~Ivanchenko, and G.~Tsironis for insightful discussions.

\end{document}